\documentclass[mathleft
]{an}
\usepackage{graphicx}
\usepackage{times}
\def\gtsima{$\; \buildrel > \over \sim \;$}
\def\ltsima{$\; \buildrel < \over \sim \;$}
\def\gtrsim{\lower.5ex\hbox{\gtsima}}
\def\lesssim{\lower.5ex\hbox{\ltsima}}

\begin{document}

\Pagespan{789}{}
\Yearpublication{2006}%
\Yearsubmission{2005}%
\Month{11}%
\Volume{999}%
\Issue{88}%

\title{Are ring galaxies the ancestors of giant low surface brightness galaxies?}

\author{M. Mapelli\inst{1}\fnmsep\thanks{Corresponding author:
  \email{mapelli@physik.unizh.ch}\newline}
\and  B. Moore\inst{1}
}
\titlerunning{Ring galaxies and GLSBs}
\authorrunning{Mapelli \&{} Moore}
\institute{
Institute for Theoretical Physics, University of Zurich,\\ Winterthurerstrasse 190, CH--8057, Z\"urich, Switzerland
}

\received{}
\accepted{}
\publonline{}

\keywords{Methods: n-body simulations -- galaxies: interactions -- galaxies: peculiar}

\abstract{%
Giant low surface brightness galaxies (GLSBs), such as Malin 1, have unusually large and flat discs. Their formation is a puzzle for cosmological simulations in the cold dark matter scenario. We suggest that GLSBs might be the final product of the dynamical evolution of collisional ring galaxies. In fact, our simulations show that, approximately 0.5-1.5 Gyr after the collision which lead to the formation of a ring galaxy, the ring keeps expanding and fades, while the disc becomes very large ($\sim{}100$ kpc) and flat. At this stage, our simulated galaxies match many properties of GLSBs (surface brightness profile, morphology, HI spectrum and rotation curve).
}

\maketitle

\section{Introduction}
The giant low surface brightness galaxies (GLSBs) are low surface brightness galaxies (LSBs) characterized by the unusually large extension of the stellar and gaseous disc (up to $\sim{}100$ kpc; Pickering et al. 1997 and references therein) and by the presence of a normal stellar bulge (Sprayberry et al. 1995; Pickering et al. 1997). Their prototype is Malin~1 (Bothun et al. 1987). The existence of GLSBs is a puzzle for cosmology and in particular for cosmological simulations in the cold dark matter scenario. In fact, most cosmological simulations including a baryonic component produce discs which are too compact and too bulge-dominated to match the properties even of a Milky Way-like galaxy (D'Onghia et al. 2006). Thus, there is no way to form GLSBs, whose discs are flat and huge, in current cosmological simulations. Various mechanisms have been proposed for the origin of GLSBs, but none of them is able to completely solve the problem. For example, a large-scale bar can redistribute the disc matter and increase the disc scalelength (Noguchi 2001). However, bar instabilities normally do not increase the disc scalelength by more than a factor of 2.5, which is not sufficient to produce the observed GLSBs.

In this proceeding, we show that the propagation of the ring in an old collisional ring galaxy can lead to the redistribution of  mass and angular momentum in both the stellar and gas component
out to a distance of $\sim{}100-150$ kpc from the centre of the galaxy, producing features (e.g. the surface brightness profile, the star formation, the HI emission spectra and the rotation curve) which are typical of GLSBs. 

\begin{figure}
\begin{center}
\includegraphics[width=75mm]{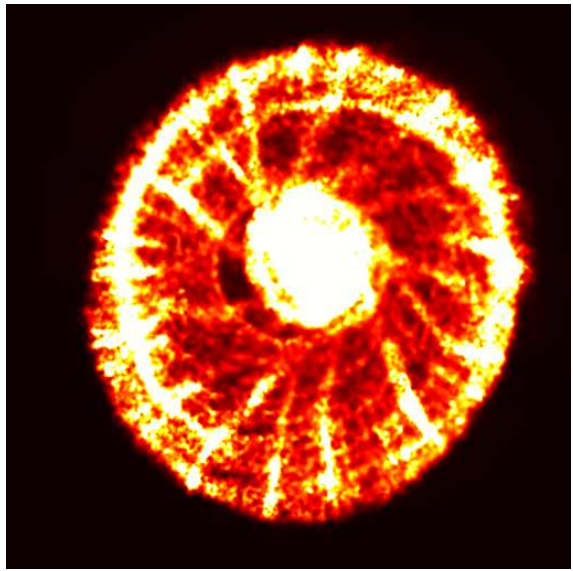}
\caption{Projected density of stars of a simulated ring galaxy. Run A3 of Mapelli et al. (2008a), $\sim{}100$ Myr after the galaxy interaction. The frame measures 90 kpc per edge. The density, projected along the $z-$ axis, scales linearly from 0 to 27 $M_\odot{}$ pc$^{-2}$.}
\label{fig1}
\end{center}
\end{figure}

\begin{figure}
\begin{center}
\includegraphics[width=80mm]{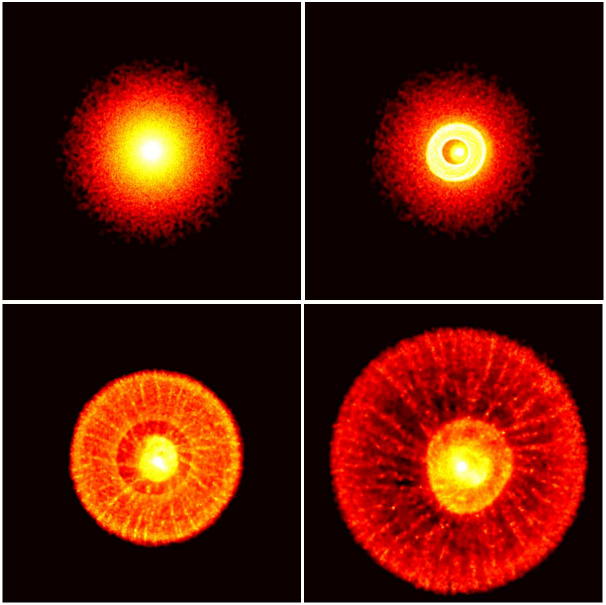}
\caption{
Projected density of gas and stars of a simulated ring galaxy. The density is projected along the $z-$ axis. Run C of Mapelli et al. (2008b). Top left panel: initial conditions. Top right panel: $\sim{}100$ Myr after the galaxy interaction. Bottom left panel: $\sim{}500$ Myr  after the galaxy interaction. Bottom right panel: $\sim{}1$ Gyr after the galaxy interaction. The frames measure 260 kpc. The colour coding indicates the density, projected along the $z$-axis, in logarithmic scale (from 0.2 to 70 $M_\odot{}$ pc$^{-2}$).}
\label{fig2}
\end{center}
\end{figure}


\section{Methods}
We simulate galaxy interactions which lead to the formation of a collisional ring galaxy. The details about the simulations are reported in Mapelli et al. (2008a) and in Mapelli et al. (2008b). Here we remind that the simulations have been done with the N-body-SPH code GASOLINE (Wadsley et al. 2004). Both the target and the intruder galaxy have a Navarro, Frenk \& White (1996, NFW) dark matter halo. The target galaxy has a stellar and gaseous exponential disc and a stellar bulge, and the ratio between the mass of the target and that of the intruder is $\sim{}2$.

\section{Results}
The target galaxy develops a well-defined ring already $\sim{}100$ Myr after the interaction (Figure~1). The morphology of the simulated ring galaxy matches many of the properties of the Cartwheel galaxy, including the `spokes' (see the appendix of Mapelli et al. 2008a). The simulation also matches the surface brightness profile of stars in Cartwheel (Higdon 1995).

We continue the simulation till $\sim{}1.5$ Gyr after the interaction. We note that the ring-galaxy phase is quite short-lived: already $\sim{}$0.5 Gyr after the interaction, the ring has propagated up to $\sim{}70-90$ kpc and its surface density has significantly lowered (Fig.~2). At the same time, the stellar disc has became extraordinarily extended and flat.
At $\gtrsim{}1$ Gyr after the encounter (bottom right panel of Figure~2) the surface density of the ring is $\sim{}2$ orders of magnitude lower than in the `Cartwheel phase', and is almost comparable with the density of the disc. The disc now extends up to  $100-130$ kpc, showing a flat surface density (in a logarithmic scale). Such flat and huge discs have been observed only in GLSBs.

\begin{figure*}
\begin{center}
\includegraphics[width=140mm]{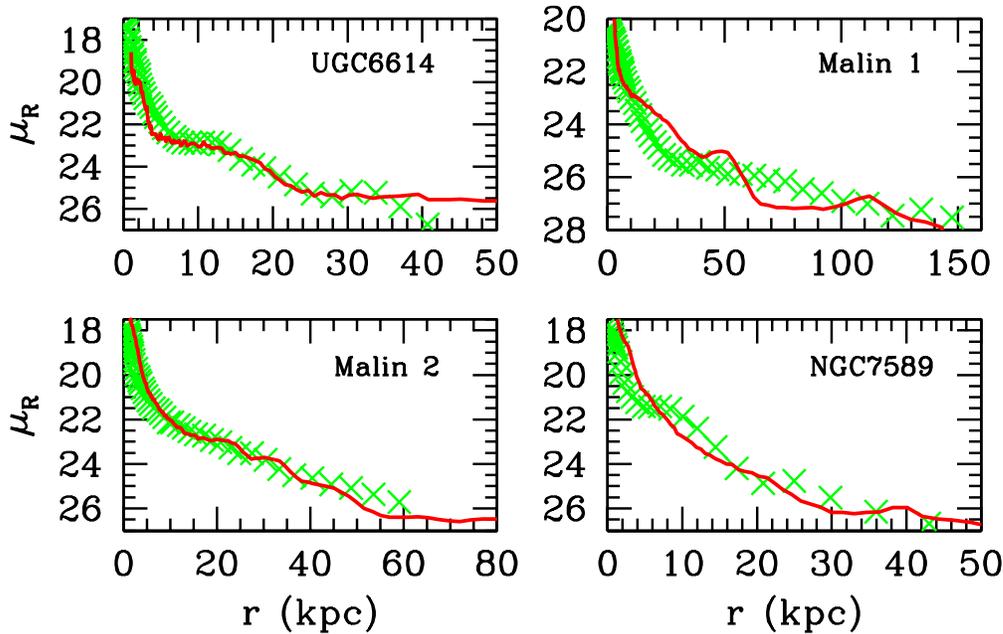}
\caption{
$R-$band stellar surface brightness profile of the GLSB sample (in units of magnitude per arcsec$^2$). Green crosses: data points from Pickering et al. (1997) for UGC~6614, Malin 2 and NGC~7589, and from Moore \& Parker (2006) for Malin 1. The $1\,{}\sigma{}$ errors are of the same order of magnitude as the points. Red solid line: stellar surface brightness profile in $R$ magnitude derived from the simulations. From top to bottom and from left to right: UGC~6614 data and run C of Mapelli et al. (2008b) at time $t=0.5$ Gyr, Malin 1 data and run C of Mapelli et al. (2008b) at $t=1.4$ Gyr, Malin 2 data and run B of Mapelli et al. (2008b) at $t=1.0$ Gyr, NGC~7589 and run A of Mapelli et al. (2008b) at $t=0.5$ Gyr.}
\label{fig3}
\end{center}
\end{figure*}

We  compare the properties of our simulated galaxies with those of the observed GLSBs. In particular, we consider a sample of four GLSBs, which have been deeply studied: UGC~6614, Malin~1, Malin~2 and NGC~7589. By means of the package TIPSY\footnote{\tt http://www-hpcc.astro.washington.edu/tools/tipsy/tipsy.html} we derive the surface brightness profiles of the simulated galaxies.
We then compare the simulated surface brightness profiles with the observed surface brightness profiles of the four galaxies. Figure~3 shows the best-matches between observations and simulations (see Mapelli et al. 2008b for details). The simulations match quite well the observations. We also derive the star formation history of the simulated galaxies, and we find a good agreement with the observed star formation rate, when available. For example, the star formation rate inferred from observations for Malin~1 is $\approx{}0.1\,{}M_\odot{}$ yr$^{-1}$, whereas the simulated star formation rate is $\sim{}0.3\,{}M_\odot{}$ yr$^{-1}$ (Impey \& Bothun 1989). The simulations also match some interesting morphological feature of many observed GLSBs, such as the existence of a bar (Pickering et al. 1997; Barth 2007).

Furthermore, we study the kinematics of the simulated galaxies. We rotate the simulated galaxy by the observed inclination angle and then we derive the velocity of gas particles along the line-of-sight, in order to produce simulated HI spectra. The results (shown in Mapelli et al. 2008b) match quite well the data by Pickering et al. (1997). With a similar technique we also derive the rotation curves of the simulated galaxy. In particular, we  rotate the simulated galaxy by the observed inclination angle, we divide the galaxy into concentric annuli and for each of them we calculate the average local velocity along the line-of-sight. The results are shown in Fig.~4. We make this kind of plots, instead than simply show the circular velocity, because we want to do something as similar as possible to what the observers do. We stress that the rotation curves that we obtain can strongly deviate from the circular velocity, as the simulated galaxies have strong non-circular motions. In fact, with the method that we adopt (and that is adopted in Pickering et al. 1997) the entire velocity along the line-of-sight is considered and there is no way of distinguishing between circular and radial motions. Thus, the rotation curves shown in Fig.~4 generally overestimate the velocity with respect to the circular velocity. This solves the apparent angular momentum discrepancy between the circular velocities of ring galaxies and of GLSBs. We also note that Pickering et al. (1997) admit the existence of strong non-circular motions in the GLSBs they analyze, and especially in  Malin~1. 
Thus, it may be important to re-analyze the existing data or to take new data and to perform a new analysis, in order to measure non-circular motions. In our simulations we have information about non-circular motions, and we can give predictions that  may be confirmed (or rejected) by future observations. In Fig.~5 the ratio between radial (v$_{\rm rad}$) and tangential (v$_{\rm tan}$) velocity is shown, as a function of radius. In all the four simulated galaxies there is a clear trend: in the inner region v$_{\rm rad}$ is negligible with respect with v$_{\rm tan}$, because the central part of the galaxy is regularly rotating. In the peripheral regions v$_{\rm rad}$ is similar to v$_{\rm tan}$, because a part of the ejected matter is rapidly falling back to the centre and the remaining is still expanding in the outer part of the ring.

\begin{figure}
\begin{center}
\includegraphics[width=80mm]{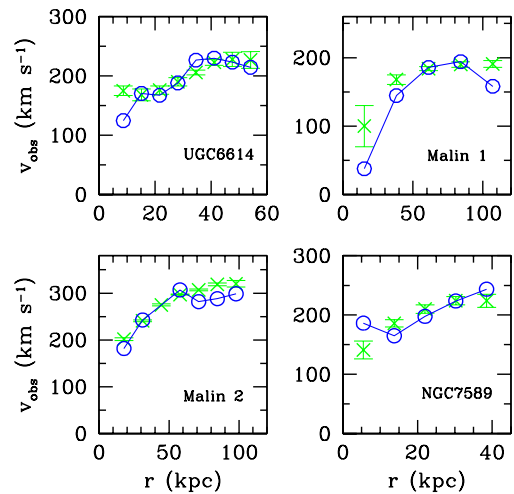}
\caption{
Rotation curves of the GLSB sample. Green crosses  are observational data from Pickering et al. (1997). $1\,{}\sigma{}$ errors are shown. Blue open circles connected by the solid line are the simulations. The simulations are the same as in Fig.~3.}
\label{fig4}
\end{center}
\end{figure}

\begin{figure}
\begin{center}
\includegraphics[width=80mm]{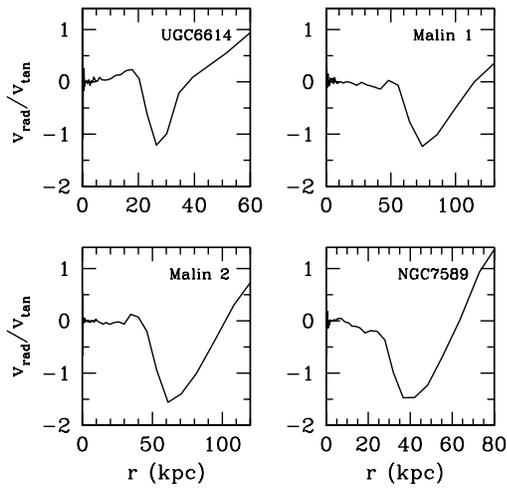}
\caption{
Ratio between radial ($v_{\rm rad}$) and tangential velocity ($v_{\rm tan}$) of gas in the simulations. The simulations are the same as in Fig.~3.
}
\label{fig5}
\end{center}
\end{figure}

\section{Discussion and conclusions}
Our simulations show that collisional ring galaxies in the late stages of their evolution ($\sim{}0.5-1.5$ Gyr) match many properties of the GLSBs (surface brightness profile, HI spectrum, rotation curves, etc.). 
This is an interesting result, as the origin of GLSBs is an open issue, so far. In particular, the proposed scenario allows to explain the origin of GLSBs within the current cosmological model.
Furthermore, the estimated local density of GLSBs ($n_{\rm GLSB}$) and that of ring galaxies ($n_{\rm RG}$) are comparable: $n_{\rm GLSB}\sim{}2.7\times{}10^{-7}\,{}h^3\,{}{\rm Mpc}^{-3}$ (where $h$ is the Hubble constant in units of 100 km s$^{-1}$ Mpc$^{-1}$) and $n_{\rm RG}\sim{}5.4\times{}10^{-6}\,{}h^3\,{}{\rm Mpc}^{-3}$ (Few \& Madore 1986). Ring galaxies are $\sim{}20$ times more numerous than GLSBs, a difference which can be explained by the fact that magnitude-limited surveys are biased against GLSBs and that a fraction of ring galaxies likely does not evolve into GLSBs.

\begin{figure}
\begin{center}
\includegraphics[width=70mm]{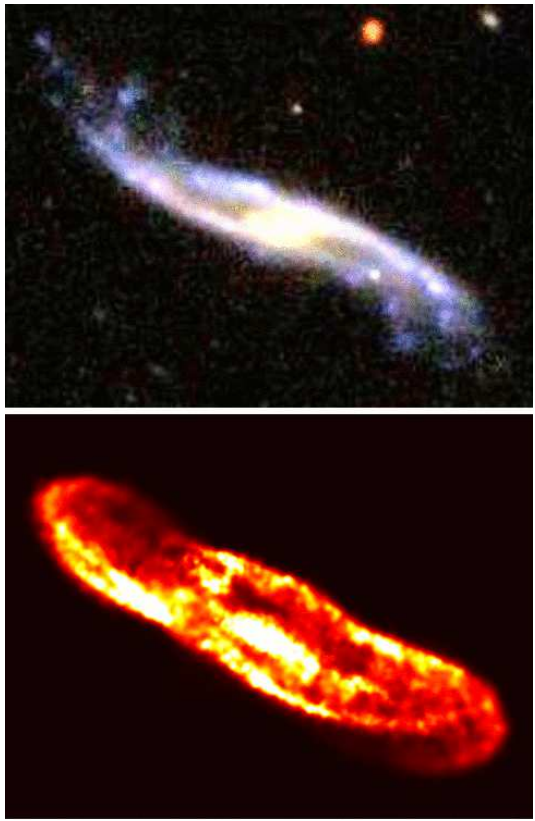}
\caption{
Top panel: SDSS composite image of UGC~7069. Bottom panel: projected density of stars in a simulation which matches the main properties of UGC~7069. See Ghosh and Mapelli (2008) for details.
}
\label{fig6}
\end{center}
\end{figure}

 Further theoretical and especially observational tests are required, in order to confirm this model of GLSB formation. First, kinematic data of GLSBs should be re-analyzed in order to find possible radial motions. For example, interesting preliminary results have been found by Coccato et al. (2007), who analyze the kinematic data of the GLSB ESO~323-G064.
Second,  it would be interesting to search for galaxies which are at the intermediate stage between ring galaxies and GLSBs. A possible example is UGC~6614, which is considered a GLSB, but shows a ring-like feature at $\sim{}10-20$ kpc from the centre. In our model this ring-like feature may be explained with the secondary ring produced by the galaxy interaction. Another interesting object is UGC~7069 (Fig.~6), which looks like a typical ring galaxy, but has a huge radius ($\sim{}50-60$ kpc, Ghosh \&{} Mapelli 2008). These peculiar objects deserve further studies.
Finally, the redshift evolution of the number of ring galaxies and of GLSBs is still an open issue. This can be addressed by means of both observations and cosmological simulations. Interesting predictions for ring galaxies have been recently presented by D'Onghia, Mapelli \&{} Moore (2008).

\acknowledgements
We acknowledge the participants of the conference 'Galactic \&{} Stellar Dynamics in the era of high resolution surveys' for the helpful discussions. MM is supported from the Swiss
National Science Foundation, project number  200020-117969/1.


\end{document}